\documentclass[12pt,preprint]{aastex}

\usepackage{natbib}

\newcommand{\um}{\,$\mu$m}

\def\aj{{AJ}}
\def\araa{{ARA\&A}}
\def\apj{{ApJ}}
\def\apjl{{ApJL}}
\def\apjs{{ApJS}}

\def\apss{{Ap\&SS}}
\def\aap{{A\&A}}

\def\mnras{{MNRAS}}

\def\pasp{{PASP}}
\def\pasj{{PASJ}}

\def\nat{{Nature}}

\def\procspie{{Proc.~SPIE}}

\slugcomment{Version \today, 
 in press at AJ} 

\shorttitle{Properties of Cold Dust in the 
Inner Disk and
Large-Scale Outflows of M\,82}
\shortauthors{Leeuw, & Robson}

\begin{document}


\title{Submillimeter Continuum Properties of Cold Dust \\ in the 
Inner Disk and
Outflows of M\,82}


\author{Lerothodi L. Leeuw\altaffilmark{1,2,3} 
and E. Ian Robson\altaffilmark{4}}

\altaffiltext{1}{Space Science and Astrophysics Branch, NASA Ames Research Center, MS 245-6, Moffett Field, CA 94035; lerothodi.l.leeuw@nasa.gov.}
\altaffiltext{2}{Department of Physics \& Electronics, Rhodes University, PO Box 94, Grahamstown 6140, South Africa; lerothodi@alum.mit.edu.}
\altaffiltext{3}{SA SKA / MeerKAT, Lonsdale Building, Lonsdale Road, Pinelands,
7405, South Africa; lerothodi.leeuw@ska.ac.za.}
\altaffiltext{4}{Astronomy Technology Centre, Royal Observatory,
Blackford Hill, Edinburgh EH9 3HJ, United Kingdom; eir@roe.ac.uk.}












\begin{abstract}
Deep submillimeter (submm) continuum imaging observations of the starburst
galaxy M\,82 are presented at 350, 450, 750 and 850\,$\mu$m
wavelengths, that were undertaken with the Submillimetre Common-User Bolometer
Array (SCUBA) on the James Clerk Maxwell Telescope in Hawaii.   
The presented maps include a co-addition of submm data 
mined from the
SCUBA Data Archive.  The co-added data produce the deepest
submm continuum maps yet of M\,82, in which low-level 850\um\
continuum has been detected out to 1.5\,kpc, at least 10\% 
farther in radius than any previously published submm  detections of
this galaxy.  The overall submm morphology and 
spatial spectral energy
distribution 
of M\,82
have a general north-south asymmetry 
consistent with H$_\alpha$ and X-ray winds, supporting the association
 of the extended 
continuum with outflows of dust grains from the 
disk into the
halo.  The new data raise interesting points
about the origin and structure of the submm 
emission in the inner disk of M\,82.  In particular,
SCUBA short wavelength evidence of 
submm continuum peaks that are
asymmetrically distributed along the galactic 
disk suggests the
inner-disk emission is re-radiation from dust concentrations along a bar
(or perhaps a spiral) rather than edges of a dust torus, as is commonly
assumed.  Higher resolution submm interferometery data 
from the
Smithsonian Submillimeter Array 
and later Atacama Large Millimeter Array 
should spatially resolve and further constrain the reported dust
emission structures in M\,82.  

\end{abstract}



\keywords{dust, extinction - galaxies: individual (M\,82) - galaxies: starburst - radiation mechanisms: thermal - submillimeter}


\section{Introduction}
\label{sec:intro_m82}

Massive ejections of gas and dust originating in galactic nuclei have
been observed at scales of a few kpc in optical emission lines,
submillimeter (submm) molecular lines, mid-infrared (MIR), ultraviolet (UV), submm and
radio continuum emission, and soft 
X-rays \citep[e.g.,][]{wat84, hec90, dev99, hoo05, eng06}.  One explanation for
the outflows is 
that a high supernova rate in the galactic nucleus heats up the
surrounding gas to high temperatures with sound speeds exceeding the
escape velocity of the galaxy, creating a $wind$ that expands outward from
the galaxy \citep{che85}.  The wind entrains cosmic rays, warm and
cool gas, as well as cool dust, making the outflow directly visible in many
wavebands.  The outflows are usually oriented along the minor axes of
the galaxies and are thus most easily observed in edge-on galaxies. 
 In local starbursts and high-z Lyman Break
galaxies with high enough global star-formation rate per unit area,
the superwinds are common and responsible for expelling metals from
these galaxies and enriching the inter-galactic medium (IGM) 
and therefore have implications in the evolution of
galaxies and the IGM
\citep[see, e.g,][for recent reviews]{hec03, vei05}.

M\,82 (NGC\,3034) is a nearby and popular object in which to
investigate the physical association between galactic nuclei and large-scale
outflows.  The galaxy is edge-on with an inclination of about
$10^{\circ}$ at position angle $72^{\circ}$ and is classified as
IrrII.  At an estimated distance of 3.63\,Mpc (as determined for M\,81
by \citet{fre94}), M\,82 has optical
dimensions of $11'.2 \times 4'.3$, i.e. $\sim 11.8 \times 4.5$\,kpc
(image scale is $\sim$17.6 arcsec$^{-1}$).  The
nuclear region, within $4' \times 2'$ about the major axis of the
galactic disk, has numerous point sources or emission concentrations,
some of which originate from supernovae and massive star clusters
and have been detected from the X-ray to radio wavebands.  Layers of dust
filaments laden these inner regions producing severe
optical extinction and copious infrared to submm re-radiated emission.


New Hubble Heritage Team optical images obtained with a deep 6-point mosaic  
in B (0.45\,$\mu$m), V (0.55\,$\mu$m), I (0.81\,$\mu$m), and H$_\alpha$
(0.65\,$\mu$m) filters  
of the Advance Camera for Surveys (ACS) on
board the {\it Hubble Space Telescope} ($HST$) exhibit detailed, filamentary
outflows of the M\,82   
\citep[][]{mut07}, especially in H$_\alpha$. 
As described above, it is thought that 
this outflow is being driven by the copious  formation of 
massive stars (or a starburst) and subsequent explosions of supernovae.
The starburst outflow not only provides the ejection mechanism for the
material from the galactic nucleus, but also heats the gas and ionizes the
hydrogen, causing it to glow with the red light of H$_\alpha$ emission line.  

These new optical images 
as well as earlier detailed natural-color 
composite images of M\,82 obtained with $HST$  
\citep[see, e.g.,]
[]{gri01} 
and the Subaru Telescope \citep[][]{ohy02}
show more than 100 compact groupings of 
about $10^{5}$ stars in very bright star clusters
sprinkled throughout the galaxy's central region, prominent dust lanes that
crisscross the disk,
knotty filaments of 
ionized gas that have rich nebular spectra that are not especially enriched in
nitrogen,
hydrogen 
gas in a strong galactic wind that
is clearly below the galactic center 
and to the right of the central region, along with many other regions of 
varying star-formation environments in the nuclear parts of this galaxy.  The
huge clusters of massive stars, numerous X-ray
and radio detected supernovae, gas concentrations, optically-dramatic
dust filaments, galactic winds and other active nuclear features
have been attributed to a large burst of star formation $10^7$ to
$10^8$ years ago, that was probably triggered by a tidal interaction
with the nearby spiral galaxy M\,81 and dwarf starburst galaxy NGC\,3077
(see, e.g., \citet[][]{yun94} for evidence of H\,I tails linking the three
and \citet
[]{sch00} for a recent review).  


The proximity of this galaxy makes it possible
to observe the region of interaction between the star-formation
regions and the halo, including expanding shells or bubbles and
``chimneys'' that
 are producing a clearer picture of the localized driving mechanisms for
the outflows \citep[e.g.,][]{hec90, wil99, wes07}.  The proximity also 
allows the detections of {\em low-level}\, emission in the halo and
consequently the determination of the amounts and possible origin of
material that results in this emission \citep[e.g.,][]{sea01,eng06}.
Studying the
contents and interactions between the star-formation regions and the
halo is important for understanding their role in the evolution of
M\,82 and may provide clues to general galaxy evolution
as well as details of the composition of the intergalactic material. 

This paper focuses on the Submillimetre Common-User Bolometer
Array (SCUBA) 
maps of the copious submm
re-radiated emission that results from the dust-laden, 
star-forming disk, as well as large-scale, low-level emission
that is associated with the outflows in the halo of M\,82. 
Submm continuum observations of the dusty central regions in M\,82 were
previously obtained with the 
submm continuum receiver UKT\,14 
on the James Clerk Maxwell Telescope (JCMT)
by \citet{hug90,hug94}
and later with SCUBA, also on the JCMT, by \citet{lee98_pro} and \citet{alt99}.
 The current study was intended to extend these previous imaging submm
observations in spatial extent, sensitivity and to all the four submm
wavelengths available with the SCUBA array  (see
Section~\ref{sec:obs_m82}).  The study includes a 
co-addition of data 
mined from the SCUBA Data Archive.  

The co-added data produce the deepest
submm continuum maps yet of M\,82, in which low-level emission is
detected out to 1.5\,kpc for the first time in 
the submm continuum of this galaxy.  The deep maps are used in a detailed
morphological study of the nuclear and large-scale detections (see
Sections~\ref{sec:general} and ~\ref{sec:m82_dis}), including a
focused comparative analysis to optical (see 
Section~\ref{sec:m82_submm_vs_opt}) and high-resolution
CO\,(1-0) (see Section~\ref{sec:m82_submm_vs_co}) morphology.  
The maps are
also used in the computation 
the first submm
spatial spectral energy distribution (SED)
of separate locations in the 
nuclear star-forming region of M\,82 (see
Section~\ref{sec:m82_dis}). 
These observational results are used in the discussion of the origin and
structure of submm continuum morphology and 
spatial SED of M\,82 and reviewed in the context of relevant interpretations
by other researchers, including those who use data from other wavelengths (see
Section~\ref{sec:m82_dis}).  In particular, (a) the commonly assumed
interpretation that the double emission peaks that are
seen in the mm to infrared continuum are due to emission from the
edges of an inclined, dusty-molecular torus is challenged (see
Section~\ref{sec:m82_origin}), (b)  
an analytical review of CO results is undertaken to assess if CO
emission may significantly contaminate the continuum observed in SCUBA
filters (see Section~\ref{sec:m82_contamination}), and (c) 
a morphological comparison is conducted to check whether the localized
outflows that are reported in high resolution radio, CO, and SiO
maps respectively by 
\citet[][]{wil99}, \citet[][]{wei99}, and \citet[][]{gar01}
can be seen in the 
SCUBA maps (see Section~\ref{sec:m82_implications}).  
  Finally,
the overall implication of the results are discussed and possible
future work is outlined (see Section~\ref{sec:m82_sum}).

\section{Observations}\label{sec:obs_m82}

SCUBA 850, 750, 450 and 350\,$\mu$m 
imaging observations of M\,82 were obtained with the telescope pointed at the
2.2\um\ infrared nuclear peak of the galaxy using positions from
\citet{die86}. 
Jiggle mapping observations were conducted with the secondary chopping in
azimuth at 7.8\,Hz and with a throw of 120\,\arcsec.
The imaging observations employed the common 64-point jiggle pattern with
a 3\,\arcsec\ offset between each position, giving fully  
sampled images with both arrays.

Because M\,82 has been a popularly
observed source with SCUBA, additional maps that were obtained with a chop
throw of 120\,\arcsec\ by other observers were mined from the
SCUBA Data Archive in order to co-add the related data and maximize
the signal-to-noise in the
final maps.  
Data sets from the 
SCUBA Archive were separately flux calibrated and corrected for
JCMT pointing errors and 
then co-added, with each observation being weighted according to its relative
integration time and the noise in the map.  The 450-850\,$\mu$m dual
mapping wavebands have been used more 
commonly than the 350-750\,$\mu$m combination, and therefore the archival
maps constitute about 85\% and 35\%  of the
respective co-added, total-integration time for the 450-850\,$\mu$m and
350-750\,$\mu$m dual maps. 

The imaging data analysis
was undertaken using the dedicated   
SCUBA data reduction software SURF \citep{jen98}, as well as KAPPA, GAIA
and CONVERT software packages provided by the Starlink
Project\footnote{The Starlink Project is run by the Council for the
Central Laboratory of the Research Councils on behalf
of the Particle Physics and Astronomy Research Council of the United
Kingdom.}.  The data reduction consisted of first flatfielding the
array images, and then correcting for atmospheric extinction.  Next, pixels
significantly noisier than the mean were blanked-out; and, after
initial inspection of raw images, pixels 
containing relatively little flux from the source were used to correct
for correlated sky noise in each individual jiggle-map.

The atmospheric opacity, $\tau$, was determined from skydips made with
SCUBA at 
intervals during the observations.  Were SCUBA-skydip measurements were
not available, the $\tau$ at the SCUBA filters was extrapolated from
the continuously measured $\tau$ 
at 225\,GHz, obtained courtesy of the Caltech Submillimeter Observatory
(CSO) radiometer and using
relations listed in Table ~\ref{tab:cso_myphd}.
At the JCMT, these relations are empirically derived  and 
periodically updated and improved as more data, especially since the
commissioning of SCUBA, have been obtained
\citep[e.g.,][]{arc02}\footnote{This latest documentation is
available at the Joint Astronomy Centre web site, 
http://www.jach.hawaii.edu}.  All the data were calibrated using instrumental 
gains determined primarily from nightly beam maps of Mars and Uranus,
or, alternatively, the JCMT secondary calibrators.  

\begin{table}[h!]
\caption[The CSO relations used in this paper]{The
CSO relations used on the data presented in this paper} \label{tab:cso_myphd} 
\vspace{2mm}
\centering
\begin{tabular}{|lll|}\hline
$ \tau_{850}$ &  = & $4.3 \times (\tau_{\rm{CSO}} - 0.007)$ \\
$ \tau_{450}$ & = & $23.9 \times (\tau_{\rm{CSO}} - 0.01)$ \\
$ \tau_{450}$ & = & $6.5 \times (\tau_{850} - 0.03)$ \\ \hline
\end{tabular}
\end{table}


\section{The General Submm Continuum Morphology}\label{sec:general}

Figure~\ref{fig:m82_850450} 
shows the 850, 750, 450 and 350\,\um\ co-added maps of M\,82 that were obtained
in 
jiggle mapping observing mode, respectively at $14\arcsec.5$,
$11\arcsec.4$, $8\arcsec.5$, and $6\arcsec.7$
resolution and about 
8, 100, 225, and 650\,mJy/beam sensitivity.
The 850 and 750\,\um\ images have a single emission peak that is centered about $9''$ west
of the galactic nucleus, while
the 450 and 350\,\um\ maps have two emission peaks centered about $10''$
and $6''$ respectively east and west 
 of the nucleus along the galactic disk.  The peak
in the west is slightly elongated along the galactic disk and is
brighter than the eastern peak, showing east-west asymmetry about the
nucleus.        
In the 750\micron\ image, the single emission peak seen in the
850\micron\ image  
begins to be
resolved-out
into the double peaks seen in the 450 and 350\micron\ images
(e.g. Figures~\ref{fig:m82_850450}, bottom panel).
This is expected, as the  750\micron\ observations have a resolution that is 
intermediate between that of the 850 and 450\micron. 
When the maps at shorter wavelengths (e.g., 350\micron) are
smoothed to resolutions similar to those at the longer wavelengths (e.g.,
850\micron), the two submm peaks that are resolved at the shorter
wavelengths become visible just as one peak, showing that indeed the
different high-brightness morphologies 
depicted in the maps are due to
the respective resolutions at 
different wavelengths.
The overall extended morphology has
an elliptical shape with  
the major axis position angle of 72$^{\circ}$, i.e. roughly the same as the
galactic disk in the nuclear region.  
  This general morphology is similar to previous continuum observations at
mm \citep[e.g.,][]{kun97,thu00}, submm \citep{hug94,lee98_pro,alt99} and
mid-infrared \citep{tel91} wavelengths, as well as to CO line transitions
\citep[e.g.,][]{nak87,thu00} 
observations of comparable resolution.


It is noted that although the maps presented in
Figure~\ref{fig:m82_850450} have similar features to
those in published submm continuum maps by \citet{hug94} and \citet{alt99}, the
very sensitive 850\,\um\ map presented here also shows that the 850\,\um\
emission (that is expected to be from cold dust) extends by at least
$10''$ ($\sim 176$\,pc) radius farther 
into the halo than detected by those authors.
 Furthermore, Figure~\ref{fig:m82_850CO2-1} 
depicts the 850\,\um\ continuum emission and integrated CO(2-1) line
 intensity maps respectively presented in Figure~\ref{fig:m82_850450} of this
 paper and Figure\,5 of work by \citet{thu00} to show that
the 850\,\um\  emission
is as extended as the CO(2-1) emission, 
contrary to claims by \citet{thu00} that the CO(2-1) emission is more
extended than the cold dust emission of M\,82, and therefore made the galaxy
exceptional in this regard.

\subsection{Submm vs. High-Resolution CO\,(1-0)
Morphology}\label{sec:m82_submm_vs_co}
 
Figure~\ref{fig:m82_850450_cocon}
 shows black contours of the
integrated-intensity-CO\,(1-0) data, that were obtained with the
Berkeley Illinois Maryland Association (BIMA) interferometer by
\citet{she95}, overlaid on the SCUBA 450 and
850\,\um\ continuum images by aligning the sky co-ordinates in the SCUBA
 maps to those of the BIMA ones.  The BIMA maps are at resolution $2''.4
\times 2''.6$ and are plotted to investigate the spatial
correspondence between the submm continuum morphology and the
high-resolution CO
features.  The top panels in Figure~\ref{fig:m82_850450_cocon} are
plotted at full resolution of the BIMA CO
contours and the bottom panels are with the CO contours smoothed to
beam sizes similar to the SCUBA-850 and -450\,$\mu$m beams ($\sim 14\arcsec.5$
and $8\arcsec.5$, respectively).
Because dust often occurs mixed-in with gas in star-forming regions, the BIMA maps are expected to give an
indication of how dust emission may appear at higher
resolution and perhaps also some insight about the structure
seen in the SCUBA images.  
The BIMA maps probably provide the best observational,
high-resolution evidence of 
structure in the inner disk that is associated with
dust, as submm continuum observations of dust in
M\,82 are not currently publicly available at a resolution higher than $6''.7$,
that is obtainable with SCUBA at 350\,\um.

The first obvious difference between the SCUBA and 
high-resolution-BIMA maps is that the dust emission peak
indicated by a red contour
is resolved into two peaks
centered about $9''$ apart in the CO maps.  Of these two CO western
peaks, the one closer to the nucleus is co-spatial with the 450\,\um\
(red contour) and slightly east of the 850\,\um\ (red contour) western
peaks.  The CO peak farthest
from the nucleus is actually the brightest of all the CO peaks and lies
on the eastern edges of the submm 850 and 450\,\um\ 
high-brightness lobes, not co-incident with the brightest submm
peaks.  
Distinct from submm continuum, the lower level CO emission near the very
western CO peak fans-out westerly at position angle
$85^{\circ}$,
diverting from the $72^{\circ}$--position-angle of
the submm continuum lobes and disk along the major axis, as well as
that of the CO within a $10''$ radius from the galactic nucleus.  
These differences in the CO and dust features are evident in both the full resolution and smoothed CO maps
plotted in Figure~\ref{fig:m82_850450_cocon}, suggesting that although the
dust and CO 
generally 
appear mixed in this star-forming region, 
in fact there are differences in their 
spatial distributions.
The 
different concentrations
are most probably 
locations of 
varying gas-to-dust densities or star-formation environments.
Consistent with the finding here of different CO-dust 
concentrations, evidence of a star-formation history
\citep[e.g.,][]{gri01} and gas density \citep[e.g.,][]{pet00} that 
clearly varies from the east to west  of the galaxy  has been reported
in M\,82.

Higher resolution submm 
observations of dust in M\,82
that should be possible with the
Smithsonian Submillimeter Array (SMA) 
and later Atacama Large Millimeter Array (ALMA)
will provide direct observational evidence
to further test how different the CO and dust emission trace each
other at the small scales shown in the BIMA maps.  Those future
observations will also test if dust
emission has more complex morphology than has currently been detected, as is
suggested by the increasing structure that is seen in the SCUBA maps
going from low to high resolutions.  Further discussion of the origin and
structure of the submm emission peaks is detailed in
Section~\ref{sec:m82_dis}.

\subsection{Submm vs. Infrared and Optical Morphology}
\label{sec:m82_submm_vs_opt}  

The spatial investigation 
of submm versus infrared and optical morphology in M\,82 is 
important 
because complex optical morphology
that is seen in 
this galaxy,
with visual 
extinction values ($A_{V}$) that range from about 3 to 25
\citep[e.g.,][]{alo01},
is thought to result from obscuration of 
optical light by large, cold dust grains that are heated by stars
(among other things) and re-radiate in the 
infrared to submm wavelengths.  
Evidence of star-formation history
\citep[e.g.,][]{gri01}, gas density \citep[e.g.,][]{pet00}, and submm
and CO emission peaks (this work, e.g. Section~\ref{sec:m82_submm_vs_co})
that 
clearly varies from the east to west  of the galaxy 
strongly suggest that any associated dust lanes must vary not
only in their geometric structure but also in their heating mechanisms
and composition.  


\subsubsection{Strong Optical-Obsuration Patches that Correspond with Submm and CO Peaks}
\label{correspond}


The 
panels in Figure~\ref{fig:m82_850450_B} 
show the  
$B$-band maps of M\,82 obtained by  
the Hubble Legacy Team 
\citep[][]{mut07}, 
overlayed with the 
450-$\mu$m continuum-emission contours shown in
Figures~\ref{fig:m82_850450}
and CO\,(1-0) 
interferometry data by \citet[][]{she95}.
The overlays were made by aligning the sky co-ordinates of the Hubble
maps with those of the SCUBA and BIMA ones.
The respective resolutions in the Hubble, SCUBA, and BIMA maps are 
$\sim 0.09\arcsec$
$\sim 8\arcsec.5$, $2\arcsec.5$. 
For best contrast, the optical intensities in 
this figure are inversely plotted, and therefore 
the light patches are strong extinction features.
The panels 
show that all the submm and CO
peaks of M\,82
are spatially coincident with very 
strong optical-obscrution patches 
seen within the intense star-forming region about $30'' \times 15''$ of the galactic nucleus 
and 2.2\um\ peak \citep[][marked by a cross in
the figures]{die86}. 
The brightest submm peak that is about $9''$ southwest
of the nucleus 
and the CO peaks in this region stretch 
along the same direction as 
prominent
optical dust lanes that lie east-west 
along the galactic disk major axis and co-spatial to these dust lanes
out to about $15''$ from the galactic nucleus. 
Interferometery maps of the central CO(1-0) peaks shown in  
the 
left panel 
of Figure~\ref{fig:m82_850450_B}
also demonstrate 
the {\em high-resolution}, spatial co-incidence between the CO(1-0) peaks and 
very dense optical-obscrution patches about the nucleus.  
The most western and brightest CO peak also co-coincides with a dense
optical-obscuration patch.

Within a radius of about $10''$ about the galactic
nucleus, or 
about two arcseconds north of the 2.2\um\ peak \citet[][]{die86},
there is intense $B-$band optical emission. 
This region is between the two submm peaks and thus
has relatively 
low submm intensity or, 
if indeed the submm
emission is from cold dust re-radiation, 
low density or heating of cold dust.  This region 
also has mid-IR  emission indicative of
star-formation clusters \citep[][]{lip04}, though of less mid-IR
brightness and lower 
mid-IR color temperature than the star-formation clusters southwest of this
position, i.e. at the location of the southwestern submm peak.
The remarkable spatial co-incidence between the submm as well as
high-resolution CO peaks and  
very dense optical-obscuration patches about the nucleus, and the
co-incidence of intense 
$B-$band optical emission
with location of relatively low submm emission or cold dust column density,
support suggestions in this paper that the submm
emission in the inner disk of M\,82 originates from re-radiation of dust
concentrations or clouds of physically different star-formation  environments
\citep[e.g.,][]{ach95,sch00,gri01}, rather than the commonly assumed
interpretation of a dusty torus about the nucleus (see
Section~\ref{sec:m82_origin} below).

In color images of M\,82 \citep[see, e.g.,][]{gri01, wes07},
the optical emission  
associated with 
submm peaks and 
high-brightness, diffuse re-radiation 
in the inner disk
of this galaxy
has a blue-brownish hue, strongly suggesting that hot, 
young, blue
stars 
are the main heating source for the dust in this inner
region.
The young stars also produce infrared emission and emission lines
associated with intense star-formation 
this galaxy \citep[e.g.,][]{ach95, alo01, lip04, wes07}.  


\subsubsection{Strong Optical-Obsuration Patches with No Corresponding Submm and CO Peaks}
\label{nocrospond}

Although all the submm and CO peaks 
correspond 
to optical features 
in the inner disk of M\,82, as described above, 
the contrary is not true; i.e,
many prominent dust, optical filaments, clouds, and
lanes have no clear corresponding submm emission counterparts.  In
particular, the submm 
emission is 
basically smooth at the locations of 
1) north-south filaments that run below, through, and flare above the
southwestern submm peak that is about $8''$ from the galactic nucleus;
2) east-west dark lanes that run west along the galactic
major axis and continue through to about $35''$ west of the galactic
nucleas (or $25''$ west of the southwestern submm peak);  
3) huge, dramatic complex of optically-obscuring clouds, filaments and
lanes that extends the entire minor axis of M\,82's disk and
covers an area greater than a
diameter of $25''$ south of the disk just east of
the eastern submm peak that is about $10''$ from the galactic
nucleus;  
and 4) light, optically-obscuring clouds and filaments in a 'low'-extinction
region known as a starburst-remnant \citep[e.g.,][]{gri01}  
that lies about $30''$ to $60''$ northeast 
of the galactic nucleus.
In another 
region of low-optical extinction 
and radiation that is about $60''$ to $120''$
northeast of the galactic nucleus,
submm emission has currently not been detected where very light
optically-obscuring filaments are evident. 

The lack of correspondence between the very dark
optical clouds, filaments, and lanes with bright submm emission suggests that
these optical features are due to obscuration by cool dust grains that
are on the near side of the galaxy and at large distances from the nuclear
region, where they are heated by a very dilute stellar radiation
field.
These foreground dust clouds evidently have enough
column densities to 
obscure optical light in the line of sight; however, they only re-emit very
low-level emission that shows no striking features in the current
submm maps.  That 
the complex optical morphology seen in
Figure~\ref{fig:m82_850450_B} 
is primarily due to foreground dust is supported by the fact that the
obscuration is more dramatic in the shorter wavelength $B-$band than
longer $I-$band images (L. L. Leeuw et al. 2009 {\it in preparation}).  Obscuring dust is expected to be 
optically thicker at the shorter wavelength and therefore cause more
optical extinction 
and thus appear more prominently at the shorter wavebands.  

One explanation for the submm low-level continuum having a relatively
smooth morphology is that, because this emission is optically thin,
the detected radiation at a particular submm wavelength represents the
total emission from the entire galactic column of dust in the line of
sight.  This is different from the optical morphology of dust because
the dust is typically seen {\em obscuring}\, stellar light, and therefore
only the dust in certain spatial stratifications (usually the
foreground) of the line-of-sight columns is observed.  In other
words, morphology due to spatial-depth or stratification of similar
dust grains that are heated by a dilute radiation field is most
often more obviously seen in optical obscuration than in submm emission.  
These morphological effects will of course depend on the sensitivities
and resolutions of
the instruments used.  In a low optical extinction and radiation region about $60''$ to $120''$
northeast of the galactic nucleus, for example, the lack of any submm
detection to date  
may simply because current continuum instruments have not been
sensitive enough       
to easily detect low-level, dust re-radition that my correspondence to
low-level extinction and stellar heating in those regions. 
The relatively larger SCUBA beam could also beam-smear and thus erase
small features detected in the higher resolution optical maps.

\subsubsection{The Outflowing Wind}
\label{outflows}

An alternative explaination for the
submm morphology of M\,82  
being smooth
as opposed to distrupted 
like the optical morphology 
depicted 
in Figure~\ref{fig:m82_850450_B} 
is that the 
submm low-level emission is primarily due to dust
entrained in {\em
outflowing}  
gas and 
physically different from the optical extinction features
that don't have any currently detected submm counterparts. 
Figure~\ref{fig:m82_850450_ha} shows the 
low- and high-brightness features of the H$_{\alpha}$ 
maps obtained by  
the Hubble Legacy Team 
\citep[][]{mut07}, overlayed with 
SCUBA 850
and 450-$\mu$m 
continuum-emission contours as shown in Figures~\ref{fig:m82_850450}.
For best contrast, the
optical intensities are inversely plotted and 
the light patches (e.g. across the center of the image in the right
panel) are foreground extinction features.   
The H$_{\alpha}$ emission is plotted saturated to high-light the
large-scale and base of the outflowing  H$_{\alpha}$ wind, so not all
the obscuration patches that criss-cross optical maps of M\,82 are
depicted here.  
The overlay in the left panel demonstrates 
the spatial co-incidence between  
the large-scale low-brightness H$_{\alpha}$ emission and 
850-$\mu$m continuum north and south of the disk, while the right panel
depicts the origin of
the large-scale H$_{\alpha}$ emission in the intense star-formation
inner disk of M\,82 about the submm emission peaks (especially near the
southwestern peak), where a
$\sim 130$\,pc expanding ``superbubble'' has been discovered in CO
\citep[][]{wei99} and ionized gas \citep[][]{wil99}. 

Recent large-scale, high-resolution OVRO observations 
by \citet[][]{wal02} 
detected, resolved molecular CO(1-0) {\em streamers} 
in and below M\,82's disk 
that have different kinematical signatures to its outflowing gas. 
Some of the streamers are well correlated with optical
obscuration features and form the basis of some prominent tidal H\,{\small I}
features \citep[][]{yun93} that are thought to provide evidence
that the gas within 
the optical disk of M\,82 is disrupted by the interaction of M\,82 with
M\,81 and likely triggers the starburst activity in M\,82's center
\citep[][]{wal02}.  
The detection of resolved  mid-IR to submm
emission that corresponds to the optical and gas streamers 
and is perhaps a physically separate component to the outflowing gas
and dust
in M\,82
should be possible
with sensitive and high-resolution mid-IR to submm imaging 
instruments using {\it Spitzer}, ALMA
and the SMA. 
New {\it Spitzer} observations reported by \citet[][]{eng06} did indeed
detect extended mid-IR emission not only in the ouflow of M\,82 but also
in its halo.  The extended halo mid-IR emission could be from material
ejected into the halo  by the outflowing wind or from the interaction of
M\,82 with M\,81 \citep[][]{eng06} . 
Future observations
with these 
sensitive instruments and their detailed data analysis 
have the potential to 1) directly 
uncover
the disruption of cold (and warm) dust distribution by the interaction of M\,82 
with M\,81, 2) decompose submm dust emission in the
disk, outflows, and streamers of M\,82 and better constrain
the properties of cold dust in these seperate components (see
Section~\ref{sec:m82_implications}), and
3) elucidate the role or consequence of the dust  
in the interaction of M\,81 and M\,82 \citep[e.g.,][]{yun93} and any
connected triggering and evolution of
the star-forming in M\,82 \citep[cf.][]{wal02}, and the re-processing of
galactic dust in general.

\section{Possible CO Contamination of the Submm
Continuum?}\label{sec:m82_contamination}  

It is worth noting that the dust morphology that is mapped in the
submm continuum from 
M\,82 may be substantially enhanced by CO emission from this galaxy.
A recent flux comparison between CO(3-2) emission and 850\,\um\
continuum in M\,82  showed that CO makes
a 47\%\, (i.e. high) contribution to the integrated continuum in
this SCUBA band \citep{sea01}.  From analyzing collated CO(4-3)
observations together with those of CO at lower 
transitions, \citet{gue93} concluded that the line
strengths in M\,82 increased as one went to higher transitions,
indicating that the higher transitions must provide 
significant  cooling in the galaxy.  All SCUBA bands have roughly the same
widths, i.e. 30\,GHz, and therefore the CO contribution to the higher frequency
continuum would be expected to be equally or more significant than that
reported for the 850\,\um\ band by \citet{sea01}.  However, because the
submm continuum in M\,82 has a thermal spectrum 
(see Section~\ref{sec:m82_fluxes}),
and therefore the submm fluxes increase with frequency, the CO
contribution to the higher frequency continuum may be less than the
estimates for the 850\,\um\ band.  For SED
analysis in this paper (e.g., Section~\ref{sec:m82_fluxes}), the CO percentage contribution to the measured
flux is assumed to be the same across
the SCUBA bands and no correction for it is made in 
the presented data. 

Observations using a new high frequency,
Fabry-Perot spectrometer on the JCMT have lead to clear detections of the
high transition CO(7-6) in M\,82 and NGC\,253 \citep{bra99}, the first
such detection in any
extragalactic sources.  The analysis by \citet{gue93} and the 
detections of CO(7-6), whose transition line lies in the 450\,\um\
filter bandpass, suggest that other
higher transition lines, such as the 13CO(8-7) line that lies at the centre of
the SCUBA 350\,\um\ band, could be very strong in M\,82, supporting the
proposition above that the high frequency SCUBA images may have
significant contribution from CO.  In this light, the morphology seen
in the SCUBA images is a direct probe of the galactic cooling and the
general interactions of active star formation and the ISM in M\,82.

Although it is not obvious if
the CO contribution to the higher frequencies of SCUBA will be less
than or as significant as estimates by \citet{sea01}, it is clear
that the CO contamination to higher-frequency continuum warrants
investigation.  Future work on
this galaxy will attempt to acquire the data of the CO lines in the
450 and 350\,\um\ bands and make a quantitative comparison of these data
in order to determine the possible contributions of CO to the
high-frequency-SCUBA data.  Such work is important (among other things) in the
determination and interpretation of submm SED
and thus the nature of dust emission in M\,82 (see
Section~\ref{sec:m82_fluxes} below). 

\section{Origin of Submm Continuum 
and 
Spectral Energy Distribution}
\label{sec:m82_dis} 

\subsection{Source and Structure of the 
Submm
Continuum in the Inner Disk}\label{sec:m82_origin}  

The radiation from M\,82 at the radio \citep[e.g.,][]{sea91, wil99}, mm \citep[e.g.,][]{hug90}
and submm-to-infrared \citep[e.g.,][]{tel91, hug94, alt99} wavelengths is respectively dominated by
synchrotron emission from supernovae,
free-free emission from ionized gas, and thermal re-radiation from dust
heated by young stars.  In this light, the double peaks
seen in the mm to infrared continuum have commonly been interpreted as due 
to emission from the edges of an inclined, dusty-molecular torus, that --
as a result of their geometry on the plane of the sky and optically-thin
nature of radiation -- have relatively high optical-depths in the line
of sight \citep[e.g.,][]{hug94}.  In Figure~\ref{fig:m82_850450}, and other
mm-to-infrared maps of similar or worse resolution (such as those from
$IRAS$), the double peaks are not resolved and appear as a single, elongated
lobe that is brightest in the southwest.  Like
the galactic disk, the lobe (or 
peaks when resolved-out) has a
position angle of roughly 72$^{\circ}$ and -- in the tori
interpretation \citep[e.g.,][]{she95} -- an inclination of $\sim 10^{\circ}$. 

The peaks of emission seen in the mm to infrared have alternatively been
interpreted simply as dust and molecular concentrations along the
galactic disk, perhaps in a bar structure \citep[e.g.,][]{nei98,wes07} that may
have an expanding ``superbubble'' of gas centered at supernova remnant 41.9+58 \citep[e.g.,][]{wei99,wil99}.  This
interpretation is supported 
by at least three reasons.  First, the east-west, double peaks have
now been seen in maps of both optically thin and thick CO emission
\citep[e.g.,][]{pet00}.  This is reasonable if the emission is from a
structure that constitutes 
concentrations or clouds of dust but is contrary to what is expected
if the emission is from a structure with tori geometry that, like the
galactic disk, is thought to be highly inclined \citep[e.g.,][]{she95}.
For optically thin radiation, it
will be possible to detect  emission from the inner parts of the
imaged structure, and either dust clouds or indeed edges of an inclined
torus would manifest as regions of relatively higher
optical-depth or brighter 
optically-thin emission in the line of sight \citep[e.g.,][]{hug94}.
However, as also noted by \citet{nei98} and \citet{pet00}, for
optically thick radiation, it will  be possible to 
directly detect emission only from the foreground surface of the
imaged structure.  In that case, the dust clouds will be seen as two
emission peaks, while the torus
(or bar) will manifest as an elongated, barlike emission of roughly
the same optical depth or optically-thick brightness.

Second, the two main peaks seen in maps of
similar resolution as the 450 and 350\,\um\ images in
Figures~\ref{fig:m82_850450} 
 are not
symmetric.  In maps of better resolution and sensitivity
\citep[e.g.,][see the CO contours in
Figure~\ref{fig:m82_850450_cocon}]{she95}, the peak 
west of the nucleus has a morphology clearly different from the
eastern peak and can be resolved into two or three structures.
High-resolution maps obtained with the Very Large Array (VLA) by \citet{wil99} showed that the western peak
is associated with
locations of supernova explosions of higher intensity and earlier
evolutionary stage than the eastern peak and confirmed the discovery
by \citet{wei99} of 
an expanding ``superbubble'' that is centered near the location of
M\,82's brightest supernova remnant, 41.9+58,
and the submm western peak.  The varying supernova intensities and ages across
the disk of M\,82 is supported by high 
resolution $HST$ imaging of stellar clusters that indicates that the
regions near the eastern and western submm peaks have different
star-formation histories \citep[e.g.,][]{gri01}.  In this light, the
submm peaks indicate concentrations of dust environments associated with 
different locations of varying supernova and star-formation
activity, not the commonly assumed dust torus.

Third, observations of line ratio gradients indicate that the average
temperature across the lobe increases from the
northeast to southwest, while the density increases in the opposite
direction \citep[e.g.,][]{pet00}.  Further evidence of the higher
temperature or, at least, column density in the southwest is seen in the
lope-sided 850 and 750\,\um\ lobe and the double-peak 450 and
350\,\um\ lobes in which the southwest parts are the brighter.  A torus
that probably houses and is heated by an active galactic nucleus
\citep[e.g.,][]{mux94}, 
would be expected to have a temperature 
that 
decreases from  
its inner to outer walls.
One explanation for the higher temperature and lower density in the 
the southwest is linked with star-formation activity that both heats
and depletes the interstellar medium (ISM) at this location
\citep{wil99}, or simply western and eastern regions of two
different star-formation physical environments
\citep[e.g.,][]{ach95,sch00,gri01,lip04}. 
  Evidence in this paper in terms of 
clearly asymmetric submm emission seen in almost all the
presented intensity
maps seems
to disfavor the commonly assumed interpretation of a dusty torus in M\,82. 

\subsection{Fluxes and 
Spectral Energy Distribution Analysis}\label{sec:m82_fluxes}

Section~\ref{sec:m82_contamination} above 
raised the possibility that CO emission may
contaminate the continuum of M\,82 observed in SCUBA filters.   For one,
significantly different contributions in the SCUBA bands
 imply different corrections to the measured fluxes 
and would affect the 
SED analysis using those
fluxes. 
If the differences are significant the SED computation and analysis
should in theory only be determined 
 after correcting or accounting for
the CO contamination.  It is  currently not obvious if the contamination to the
higher-frequency continuum will be less or higher than the {\em 47\%}
estimated to the SCUBA-850\,\um\ band by
\citet{sea01}; even though it is clear that it may also be important (see
Section~\ref{sec:m82_contamination}).  While relevant CO data need to 
be acquired in the future to make a quantitative calculation of the relative
contributions in the SCUBA bands, for the practical determination of
the 
SEDs of M\,82 in this paper, it is assumed 
that the CO contribution in 
the SCUBA bands is the same.  

\begin{table}[h]
\centering
\caption[]{
\small{
SCUBA continuum fluxes for specific locations in 
M\,82
(The listed measured flux
   densities 
include roughly 47\% possible contribution from C0 as discussed in the text.)
}}\label{tab:fluxes}
\begin{tabular}{lcccc}
\hline
\multicolumn{1}{c}{Locations parallel to M\,82's} & \multicolumn{1}{c}{Flux (Jy)} & \multicolumn{1}{c}{Flux (Jy)} & \multicolumn{1}{c}{Flux (Jy)} & \multicolumn{1}{c}{Flux (Jy)} \\
disk position angle of 72$^{\circ}$ & @350$\mu$m & @450$\mu$m & @750$\mu$m & @850$\mu$m \\
\hline
@Peak Flux  & $  18.6 \pm 5.6 $ & $ 13.0 \pm 2.6 $ & $ 1.8 \pm 0.4 $ & $
 1.2 \pm 0.1 $  \\
$30'' \times 15''$, about nucleus & $ 49.3 \pm 13.7 $ & $ 28.8 \pm 5.6 $ & $ 3.2
 \pm 0.5 $ & $  2.3 \pm 0.2 $  \\
$70'' \times 40''$, about nucleus & $ 63.4 \pm 18.9 $ & $ 35.9 \pm 7.0 $ & $ 7.7
 \pm 1.5 $ & $  3.8 \pm 0.4 $  \\
$30'' \times 15''$, $15''$N of nucleus & $  15.9 \pm 4.8 $ & $ 7.6 \pm
 1.5 $ & $ 1.0 \pm 0.2 $ & $  0.6 \pm 0.1 $  \\
$30'' \times 15''$, $15''$S of nucleus & $ 12.8 \pm 3.8 $ & $ 8.1 \pm
 1.6 $ & $ 0.7 \pm 0.1 $ & $ 0.6 \pm 0.1 $  \\
$30'' \times 15''$, $30''$N of nucleus & $  6.2 \pm 1.9 $ & $ 2.8 \pm
 0.5 $ & $ 0.3 \pm 0.1 $ & $  0.2 \pm 0.02 $  \\
$30'' \times 15''$, $30''$S of nucleus & $ 3.1 \pm 0.9 $ & $ 1.8 \pm
 0.4 $ & $ 0.2 \pm 0.1 $ & $ 0.1 \pm 0.1 $  \\
\hline
\end{tabular}
\end{table}


Table~\ref{tab:fluxes} shows the submm fluxes measured at specific
locations across M\,82 using SCUBA.  The listed errors include
calibtation uncertainties.  The submm dominant emitting region in
M\,82 is within $30'' \times 15''$ about the nucleus and is associated
with the most intense star-formation in the galaxy.  Assuming the primary
source of the submm emission  
is dust re-radiation, the 
measured fluxes from this galaxy 
were fitted with the following thermal function: 
\vspace{-0.15in}
\begin{equation}
F_{\nu} = \Omega B_{\nu}(T)
[1-{\rm exp}(-({\frac{\lambda_o}{\lambda}})^{\beta})],
\label{eqn:pb}
\end{equation}
where $\Omega$ is the solid angle for the emitting region,
$B_{\nu}(T)$ the Planck function at
temperature $T$, $\lambda_o$ the wavelength at which the optical depth
is unity ($\lambda_o = 7.8\,\mu$m, \citet[][]{hug94}), and $\beta$ the emissivity index of the grains.  
Due to the limited
frequency sampling in the data, the temperature $T$ and emissivity index
$\beta$ were
the only parameters that were statistically determined in
Equation~\ref{eqn:pb}.

\begin{table}[h]
\centering
\caption[]{
\small{
SCUBA derived 
dust emission properties
for 
locations in 
M\,82 
}}\label{tab:temp}
\begin{tabular}{lccc}
\hline
\multicolumn{1}{c}{Locations parallel to M\,82's} & \multicolumn{1}{c}{$T$}
 & \multicolumn{1}{c}{$\beta$} & \multicolumn{1}{c}{$\Omega$} \\
disk position angle of 72$^{\circ}$ & (K) & & (sr) \\
\hline
$30'' \times 15''$, about nucleus & $ 36 \pm 14 $ & $ 2.0 \pm 0.1 $ & $ 1.66e-8
 $ \\
$30'' \times 15''$, $15''$N of nucleus & $  31 \pm 3 $ & $ 2.2 \pm
 0.1 $ & $ 1.66e-8 $ \\
$30'' \times 15''$, $15''$S of nucleus & $ 31 \pm 15 $ & $ 2.3 \pm
 0.2 $ & $ 1.66e-8 $ \\
$70'' \times 40''$, about nucleus & $ 27 \pm 5 $ & $ 1.8 \pm 0.1 $ & $ 9.05e-8 $ \\
\hline
\end{tabular}
\end{table}

Table~\ref{tab:temp} lists the
derived dust $T$ and $\beta$, as well as the $\Omega$ associated
with the specified emitting regions in M\,82.
For the 
listed rectangular locations, the $\Omega$ is
determined from the longest side of the rectangle.  The average derived $T$ is
$\sim 31$\,K, with a minimum and maximum of $\sim 27$\,K and $\sim
36$\,K respectively. The highest $T$'s are from regions of
the highest surface brightnesses, 
presumably corresponding to regions of intense
star-formation, while the coldest $T$'s are at regions farthest from the
nucleus. The $\beta$ values range from 1.8 to 2.3 
and are highest at regions farthest from the
galactic nucleus.  The higher $\beta$ values are associated with larger and
colder dust grains; therefore, the spatial SED analysis here points to colder
grains more prevelant with increasing distance
from the galactic center of M\,82, particularly 
along the minor axis of the galactic disc (see
Section~\ref{sec:m82_implications} below).   

If the CO contamination in the SCUBA bands were not constant for M\,82, as assumed
in this paper, an SED with SCUBA fluxes may have looked different than
the result above and possibly required a different interpretation. For
example, a CO contamination that increased with frequency, as suggested
from an analysis of lower-transition CO and
CO(4-3) data of M\,82 by \citet[][]{gue93} (see
Section~\ref{sec:m82_contamination}), would mean that SCUBA fluxes 
uncorrected for CO contamination lead to the SED indicating lower
dust temperatures and/or higher emissivity indices than was actually the
case.  A proper correction for any CO contamination is therefore needed
before a more definitive SCUBA SED analysis can be conducted for M\,82
and is worth attempting in the future, when 
CO data in all the SCUBA
wavebands (especially the 350 and $450\mu$m ones) are available.



\subsection{Implications of the SCUBA data and SEDs on the Outflow of Cold
Dust}\label{sec:m82_implications}   


In has been noted that the
continuum morphology in the inner-halo of M\,82
has a general north-south asymmetry, that is consistent with the 
north-south asymmetric X-ray and H$_{\alpha}$ winds
\citep[e.g.,][and see Section~\ref{outflows}]{wat84} and the associated
UV, optical, molecular and indeed infrared to mm structures that have been
reported in M\,82 \citep[e.g.,][]{sea01}.  Therefore, a simple
interpretation of the asymmetric
morphology of the submm continuum in the halo of M\,82 and is that 
it is a manifestation of an outflow of dust from the 
inner disk to the halo.

\citet[][]{sea91} presented one of the earlier, extensive evidence of the
disk-to-halo outflows from spectral index distribution computations
using radio continuum maps at several wavelengths between 0.33 and
4.9\,GHz (90 and 6\,cm).  They found spectral indices between --0.3 and
--0.6 in the 
inner-disk region, steepening to about --1.0 at a radius of
about 1\,kpc along the minor axis, and concluded that these were from
relativistic synchrotron-emitting electrons that were being scattered
against infrared photons emitted in the 
inner-disk region 
of M\,82.  Recently,
\citet{wil99} used high resolution VLA continuum data between 1.4 and
5\,GHz and computed spectral indices from $-0.6$ to $-0.8$ about
$20''$ north of the 
disk, in localized nuclear sites of the outlows
that they call `chimneys'.   These values are consistent
with the results by \citet[][]{sea91} in the same wavebands, and \citet{wil99}
also interpreted them as indicating synchrotron emission from relativistic
electrons entrained in the wind.

A thermal component in the filaments has previously been  
suggested based on `tentacles' observed in Ne\,{\small II} maps of
M\,82 \citep{ach95}, which presumably are also a manifestation of the
outflow phenomenon.   
In comparative SED analysis of $30'' \times 15''$ regions centered in galactic
nucleus and two others $15''$ north and south of it, the regions north
and south had respectively cooler temperatures and higher emmisivity
indices than the central region.  This spatial SED analysis is consistent with
the submm emission coming from 
a thermal source with a temperature decrease and emmisitivity
index increase along the minor axes of the disk of M\,82.  
The change of the dust properties along the minor axes 
has a similar direction 
to the radio spectral index gradient shown by \citet[][]{sea91} and
\citet[][]{wil99} and is consistent with the north-south 
asymmetric
H$_{\alpha}$ winds \citep[e.g.,][]{sho98} of M\,82 that has 
been shown to be co-spatial
 with the submm morphology (see Figure~\ref{fig:m82_850450_ha} and Section~\ref{outflows}).  In
this light, the submm continuum maps (and SED changes along the minor
axes) indicate an outflow of dust
grains that are ejected from the 
inner disk by, or entrained in,
the starburst winds.   

One explanation for the outflow being asymmetric was given by
\citet{sho98}, who suggested that if the star-forming
disk is slightly shifted up from the galactic disk, this would imply 
that there is less covering material in the north and would make
collimation difficult, resulting in an immediate blow-out of material
in the north.  Detections confirming that collimation is better to the
south of M\,82 have been made of large-scale emission extending to 1.5\,kpc,
and more extended in the south,
not only in optical line maps \citep[e.g.,][]{dev99}, but also in
CO(2-1) and  CO(3-2) respectively by \citet{thu00} and \citet{sea01}.
Another valuable result of the co-addition of SCUBA archive data
in this paper is that the most
sensitive submm maps that are displayed in Figure~\ref{fig:m82_850450}
show submm extended emission that is associated with the outflows on
scales that for the first time match the 1.5\,kpc CO detections noted above.

Recently reported radio `chimneys'
\citep[e.g.,][]{wil99}, that are about $20''$
north of the disk and hypothesised to signify local blow-outs of material by
supernova-driven winds, are not obvious in the SCUBA maps.   However, 
prominent SiO features associated with the localized radio outflows have now
been detected in mm-heterodyne observations obtained with the Institut
de Radio Astronomie Millimetrique
interferometer \citep[e.g.,][]{gar01}.  These authors
explain the SiO detections in a framework of shocked chemistry at the sites 
of the gas ejections from the starburst disk.  For the moment it
appears that the shocked gas has proved to be a better probe of the localized
outlows than the direct observations of dust emission in the mm to
submm continuum. 

Localized sites of the outlows or `emission spurs', although long
sought-after and sometimes reported in mm to submm continuum and CO maps
\citep[e.g.,][]{hug94,she95,lee98_pro,alt99},
have not been reliably reproduced in the different mm to submm
observations.  All the continuum maps in
this paper also have some low-level `spurs', but almost none are
reproduced at exactly the same locations and extend to the same
degrees between any two different observations.  This would suggest that
the spurs might be
artefacts in the maps.  However, if the mm and submm spurs are emission from 
dust outflows 
(and filaments, clouds or lanes) that 
are of different optical 
depths or compositions and re-radiate low-level emission in relatively narrow
wavebands, the submm spurs may not be reproduced in maps at certain wavelengths
and sensitivities.  

It was noted in Section~\ref{sec:m82_submm_vs_opt} that dust
in M\,82 is most probably within components of physically varying 
locations and origin or simply at different
spatial-depths or stratifications.
The recent high-resolution,
interferometry observations of M\,82's molecular gas  by \citet[][]{wal02}
discovered  CO(1-0) `streamers' and decoupled previously observed
CO outflows \citep[e.g.,][]{nak87,sea01} from the streamers \citep[e.g.,][]{yun93}, 
clarifying the spatial distribution and origin of the molecular gas in
this galaxy.   
Similar, future observations
with recently commissioned 
sensitive and high-resolution submm imaging 
instruments such as {\it Spitzer}
and the SMA should provide tighter constraints on the spatial and optical depth
properties of dust, 
test the reality of the 
reported submm
spurs, and possibly detect dust re-radiation 
streamers, clarifying the
implications or associations of all these features to dust 
outflows and 
recycling in M\,82 (c.f. Section~\ref{sec:m82_submm_vs_opt}). 


\section{Summary of Results and Future Work on M\,82}\label{sec:m82_sum}  

SCUBA 350, 450, 750 and 850\,$\mu$m imaging observations have been
presented of the dust-laden, star-forming 
inner disk and
large-scale, low-level emission that is associated with the outflows
in the halo of M\,82.  The displayed maps include co-added data that
were mined from the SCUBA  Data Archive, resulting in the deepest
submm continuum maps of M\,82.  The 850\,\um\ morphology has a single
emission peak that is centered about $9''$ west  
of the galactic nucleus, while
the 450 and 350\,\um\ maps have two emission peaks centered about $10''$
and $6''$ respectively east and west  of the nucleus along the
galactic disk, similar to previous continuum 
observations at mm, submm and mid-infrared wavelengths, as well as to
CO line transitions and H$_{\alpha}$ observations of comparable
resolution (see Section~\ref{sec:general}).  In the 750\um\ image, the single emission peak
seen in the 850\micron\ image (see Figure~\ref{fig:m82_850450} 
is predictably beginning to be resolved-out
into the double peaks seen in the 450 and 350\micron\ images.  Low-level emission is 
detected out to 1.5\,kpc for the first time in the 850\,\um\ continuum of
this galaxy, i.e. at least $\sim 160$\,pc radius farther-out than other recent
studies.  

The deep maps were used
in a detailed morphological study of the 
disk and large-scale
detections, including a comparative analysis of submm to optical
morphology (see 
Section~\ref{sec:m82_submm_vs_opt}).
The overall, extended submm morphology of M\,82 generally resembles
the optical picture in that the 
disk emission has an apparently elliptical
shape whose major axis is clearly aligned with that of the galactic
disk at position angle $\sim 72^{\circ}$.  However, the submm
morphology is much smoother than the optical picture, and some
prominent dust cloud and filamentary lanes that are seen in the optical  
are not obvious in the submm continuum.  
One simple 
explanation for 
the submm versus optical correspondence (or lack of it) 
is the different resolutions and sensitivies 
of the presented observations.  
If the
optical features emit submm emission, it is possible the emission is 
at a level lower than
the current mapped SCUBA sensitivities 
or smeared and thus erased by the relatively larger SCUBA beam. 
This can be 
varified by future higher resolution and more sensitive
submm observations 
that should become possible with ALMA.

A comparative analysis was further conducted of submm to high-resolution
CO\,(1-0) morphology (see Section~\ref{sec:m82_submm_vs_co}).
Some resolved peaks in the CO maps could be associated with unresolved
features in the submm maps.  
However, there are differences 
that show the CO and dust emission do not trace each other in a very
simple way and 
suggest that although the dust and CO 
generally  appear mixed in the central star-forming region, 
in fact there are differences in their 
spatial distributions.
The 
different concentrations
are most probably 
locations of 
varying gas-to-dust densities or star-formation environments, 
that have been reported in M\,82 \citep[e.g.,][]{gri01,pet00}.
This can be 
varified by future higher resolution
observations of dust in M\,82
that should be possible with the
SMA and later ALMA.

The SCUBA maps were also
 used in a computation of the first submm 
spatial SED analysis of locations within and outside
the central star-forming region
of M\,82 (see Section~\ref{sec:m82_fluxes}) and in
the discussion of the origin and structure of submm maps (see
Section~\ref{sec:m82_dis}).  In particular, (a) the commonly assumed
interpretation that the double emission peaks that were
seen in the mm to infrared continuum are due to emission from the
edges of an inclined, dusty-molecular torus was challenged (see
Section~\ref{sec:m82_origin}), (b)  
an analytical review of CO results was undertaken to assess if CO
emission might significantly contaminate the continuum observed in SCUBA
filters (see Section~\ref{sec:m82_contamination}), and (c) 
a morphological comparison was conducted to check whether the localized
outflows that were reported in radio and SiO maps respectively by
\citet[][]{wil99} and \citet[][]{gar01} could be seen in the 
SCUBA maps (see Section~\ref{sec:m82_implications}).  

Evidence in this paper in terms of 
clearly asymmetric 
inner-disk submm emission seen in 
the presented intensity 
seems
to disfavor the commonly assumed interpretation of a dusty torus in M\,82. 
Arguments were presented to explane the 
inner-disk submm maps
of M\,82 in the context of emission from a rather complex
distribution of dust concentrations that are in regions of
different star-formation environments, as has been reported from
various studies using data at other wavelengths
\citep[e.g.,][]{ach95,sch00,gri01}.  

It is not obvious if
the CO contribution to the higher frequencies of SCUBA will be less
than or higher than the 47\% estimated to the SCUBA-850\,\um\ band by
\citet{sea01}.  However, it is clear
that the CO contamination to higher-frequency continuum may also be
significant and warrants detailed investigation.  Future work on
this galaxy will attempt to acquire the data of the CO lines in the
450 and 350\,\um\ bands and make a quantitative comparison of these data
in order to determine the possible contributions of CO to the
high-frequency-SCUBA data.    

The overall submm low-level morphology
has a general north-south asymmetry that is similar to the
H$_\alpha$ winds and CO and X-ray outflows that have been
detected in M\,82 \citep[e.g.,][]{sho98}.  The submm
spatial SED analysis also shows thermal properties that change along the
minor axis of the galaxy disk,
similar to
the radio spectral index gradient by \citet[][]{sea91} and
\citet[][]{wil99} and consistent with the north-south asymmetric,
large-scale X-ray and H$_{\alpha}$ winds.  Therefore, the current
results support the simple interpretation
\citep[e.g.,][]{lee98_pro,alt99} that the asymmetric
morphology in the submm maps is a manifestation of corresponding 
outflows of dust grains from the galactic 
disk into the halo.
As noted above, this work has presented low-level 850\,\um\ continuum
emission out to 1.5\,kpc for the first time in this galaxy, i.e. at
least $\sim 160$\,pc radius farther-out than other recent studies.



\acknowledgments


\section{Acknowledgments}

Lerothodi L. Leeuw acknowledges a NASA Postdoctoral Fellowship that
supported the final write-up of this work at NASA
Ames Research Center.

\clearpage

\begin{figure}
\begin{centering}
\caption[Deep maps of M\,82 at 850, 750, 450, and 350\,$\mu$m]
{Deep maps of the 850 {\it (top left - a)}, 750 {\it (top
 right - b)}, 450 {\it (bottom left - c)}, and 350\,$\mu$m {\it (bottom
 right - d)}
continuum emission
centered at the near-infrared nucleus of M\,82. 
The 850, 750, 450, and 350\,$\mu$m contours on the respective maps 
are [16, 32, 48, 64, 96, 128, 200, 300, 400, 600, 800, 1000, 1175, and
1350], [200, 400, 600, 800, 1000, 1300, 1600, and
1900], 
[450, 900, 1450, 2000, 3000, 4000, 5000, 6000, and 6800], and 
[1300, 1950, 3250, 5200, 7000, 8500, and 10000]\,mJy/beam; 
the respectively rms uncertainties are 
$\sim 8$, 100,
225, and 650\,mJy/beam, and the resolutions are 
$\sim 14\arcsec.5$,
$11\arcsec.4$, $8\arcsec.5$, and $6\arcsec.7$.
The white circle in each map indicates the approximate size
 of the SCUBA beam at the plotted wavelength.  The keys are
grayscale-coded intensities in Jy/beam, and the X and Y axes are
J2000 coordinates.  \label{fig:m82_850450}}
\end{centering}
\end{figure}


\clearpage

\begin{figure}[h]
\begin{centering}
\caption[Deep maps of M\,82 at 850, 750, 450, and 350\,$\mu$m]
{Deep maps of the 850\,$\mu$m continuum emission {\it (left - a)} and integrated CO(2-1) line
 intensity {\it (right - b)} 
of M\,82 respectively as presented in Figure~\ref{fig:m82_850450} of this
 paper and Figure\,5 of work by \citet{thu00}.  
The respective beams for the two observations are plotted and the rest
 of the keys, contour lines, axes are as presented in the indicated
 figures. A scale bar of $\sim 1.5$\,kpc is shown on the maps. 
\label{fig:m82_850CO2-1}}
\end{centering}
\end{figure}

\clearpage

\begin{figure}
\begin{centering}
\caption
[The central $76'' \times 56''$ region of M\,82 at 850 and
450\,$\mu$m overlaid with high-resolution CO\,(1-0) contours]{The central $76''
\times 56''$ region of M\,82 
at 850 (left panels - a \& c) and 450\,$\mu$m (right panels - b \& d)
overlaid with the integrated-intensity (black contours) of the
CO\,(1-0) interferometry data by 
\citet[][]{she95}.  The top panels are plotted at full resolution of the CO
contours and the bottom panels are with the CO contours smoothed to
beam sizes similar to the SCUBA-850 and -450\,$\mu$m beams ($\sim 14\arcsec.5$
and $8\arcsec.5$, respectively).
The white
850 and 450\,$\mu$m contours on the 
respective maps  are [200
to 1175] and [900 to
 6000]\,mJy/beam, as displayed in Figure~\ref{fig:m82_850450}.  The red 850 and
450\,$\mu$m contours are at 1350 and  6800\,mJy/beam and are plotted
to highlight the position of the respective western submm peak.  The
cross marks the 2.2\um\ peak \citep[][]{die86}.  The keys are
 color-coded intensities in Jy/beam and
the X and Y axes are J2000 coordinates.\label{fig:m82_850450_cocon}} 
\end{centering}
\end{figure}

\clearpage

\begin{figure}[h]
\begin{centering}
\hspace{-.6cm}
\vspace{-0.2cm}
\caption[Contours of M\,82 from the 850 and 450\,$\mu$m continuum
obtained with a 120\arcsec\ chop throw]
{The left (a) and right (b) panels respectively show 
the B-band maps of M\,82 obtained by  
the Hubble Legacy Team 
\citep[][]{mut07}, respectively overlayed with 
the 
450-$\mu$m continuum-emission contours shown in
Figures~\ref{fig:m82_850450} and
CO\,(1-0) 
interferometry data by \citet[][]{she95}.
For best contrast, the
optical intensities are inversely plotted and 
therefore the light patches are extinction features.
The centeral cross and open squares respectively mark positions of  
the 2.2\um\ peak by \citet[][]{die86} and mid-infrared,
star-forming clusters by \citet[][]{lip04}.  
The X and Y axes are J2000
coordinates.\label{fig:m82_850450_B}} 
\end{centering}
\end{figure}

\clearpage

\begin{figure}[h]
\begin{centering}
\vspace{-.1cm}
\caption
[Contours of M\,82 from the 850 and 450\,$\mu$m continuum on the H alpha]
{The left (a) and right (b) panels respectively show the 
low- and high-brightness features of the H$_{\alpha}$ 
maps obtained by  
the Hubble Legacy Team 
\citep[][]{mut07}, overlayed with 
SCUBA 850
and 450-$\mu$m 
continuum-emission contours as shown in
Figures~\ref{fig:m82_850450}.
The overlays were made by aligning the sky co-ordinates of the Hubble
maps with those of the SCUBA ones.
The respective resolutions in the Hubble, SCUBA 850-$\mu$m, and 450-$\mu$m  maps are $\lesssim
1\arcsec$, 
$\sim 14\arcsec.5$, and 
$\sim 8\arcsec.5$.
The overlay in the left panel demonstrates 
the spatial co-incidence between  
the large-scale low-brightness H$_{\alpha}$ emission and 
850-$\mu$m continuum, while the right panel depicts the origin of
the large-scale H$_{\alpha}$ emission in the intensive star-formation
inner region of M\,82's disk about the submm emission peaks.
For best contrast, the
optical intensities are inversely plotted and 
the light patches (e.g. across the center of the image in the right panel) are foreground extinction features.  The X and Y axes are J2000
coordinates.\label{fig:m82_850450_ha}} 
\end{centering}
\end{figure}

\end{document}